\documentclass[twocolumn,showpacs,aps]{revtex4}
\usepackage{bm}

\begin{document}
\title{Local deterministic model of singlet state correlations}

\author{Michael J.W. Hall}
\affiliation{Theoretical Physics, Research School of Physics and Engineering,  Australian National
University, Canberra ACT 0200, Australia}

\begin{abstract}
The derivation of Bell inequalities requires an assumption of measurement independence, related to the amount of free will experimenters have in choosing measurement settings.  Violation of these inequalities by  singlet state correlations, as has been experimentally observed, brings this assumption into question.  A simple measure of the degree of measurement independence is defined for correlation models, and it is shown that all spin correlations of a singlet state can be modeled via giving up a fraction of just 14\% of measurement independence. The underlying model is deterministic and no-signalling. It may thus be favourably compared with other underlying models of the singlet state, which require maximum indeterminism or maximum signalling.  A local deterministic model is also given that achieves the maximum possible violation of the well known Bell-CHSH inequality, at a cost of only $1/3$ of measurement independence.
\end{abstract}

\maketitle

{\it Introduction:}
One of the most remarkable features of quantum mechanics is that its predictions violate certain statistical inequalities, called Bell inequalities.  These inequalities can be derived from various sets of very plausible physical postulates, and thus their violation raises very deep issues:  either quantum mechanics makes incorrect predictions, or at least one of the postulates used is inapplicable in the description of natural phenomena.  Since all tests of quantum mechanics have so far passed experimental scrutiny, it is therefore of great interest to analyse the assumptions used in the derivation of Bell inequalities, and in particular the degree to which these assumptions must be relaxed to model quantum systems.

One critical assumption made in the derivation of Bell inequalities is measurement independence:  that measurement settings can be chosen independently of any underlying variables describing the system.  Measurement independence has not been given much serious attention in the literature - the postulate that experimenters can freely choose between measurement settings is generally explicitly acknowledged, but rarely questioned.  Shimony et al. have emphasised the reasonableness of this postulate via an amusing scenario in which two physicists, their secretaries, and the experimental apparatus manufacturer unconsciously `conspire' to choose just the right measurement settings to violate the Bell inequalites, in a fully local and deterministic manner \cite{shimony}.   
However, reasonableness alone is not sufficient: other typical assumptions made in derivations of Bell inequalities, such as no-signalling, factorizability or determinism, are also very reasonable.  It is therefore important that {\it all} assumptions leading to Bell inequalities are critically investigated - including measurement independence in particular.

In this regard, the main existing result is due to Brans, who gave an explicit local and deterministic model for correlations between any two spin-1/2 particles \cite{brans} (as well as an excellent discussion of the assumption of measurement independence in the prior literature). In this model, an underlying random variable fully determines not only the joint measurement outcomes, but also the associated measurement settings - i.e., there is no measurement independence at all.  Brans points out that this might even be considered a reasonable property of any fully causal model underlying nature - the detector settings should be predicted by such a model to the same extent as the measurement outcomes, with no {\it a priori} reason why one should not be correlated with the other.

Here it will be shown that one in fact does not have to give up measurement independence completely, as per Brans (or Shimony et al.).  Indeed, introducing a suitable measure of the degree of measurement independence, one needs to relax this degree by only 14\%, to obtain a no-signalling and deterministic model of the singlet state.  This model can therefore be said to allow a high degree of `experimental free will'.  It will also be shown that the well known Bell-CHSH inequality can be maximally violated by a local deterministic model, while allowing $2/3$ of the maximum possible measurement independence.  Here, `deterministic' indicates that the values of measurement outcomes are fully specified by an underlying variable, which is averaged over to generate the joint probabilities being modeled.

The results complement those of Branciard et al., who have shown that any model of the singlet state which satisfies measurement independence and no-signalling must be maximally indeterministic (i.e., with marginal spin probabilities of 1/2 in all circumstances) \cite{branc}.  That is, determinism must be given up completely in this case.  Further, Toner and Bacon have instead relaxed the assumption of no-signalling, and have shown there is a model of the singlet state satisfying both measurement independence and determinism, but which requires a full bit of nonlocal signalling \cite{toner}. 

Thus, perhaps surprisingly, relaxing the assumption of measurement independence may be favourably compared to relaxing either determinism or no-signalling: only 14\% needs to be given up in the former case, in comparison to either 100\% of determinism or 100\% of no-signalling, to model the singlet state. 

Note that Conway and Kochen have derived a theorem implying that particular types of correlations cannot be modelled under assumptions of measurement independence, no-signalling and determinism \cite{conway}.  In this sense their theorem is similar in significance to the derivation of Bell inequalities.  However, they do not consider relaxing any of these assumptions.  Further, while they appear to equate both measurement independence {\it and} indeterminism with `free will' (for experimenters and particles respectively), their conclusion that particles `have exactly the same kind' of free will as experimenters is somewhat unclear.  Indeed, it will be shown elsewhere that the particular correlations considered by Conway and Kochen 
have a no-signalling model which has 95\% measurement independence but 0\% indeterminism  \cite{relaxed}. 

{\it Quantifying measurement independence:}
Any underlying model for a set of joint probabilities  $p_{XY}(a,b)$, corresponding to the probabilities of obtaining respective outcomes $a$ and $b$ for joint measurement settings $X$ and $Y$, postulates the existence of some underlying variable $\lambda$, such that 
\begin{equation} \label{underlying}
p_{XY}(a,b) = \int d\lambda \, \rho_{XY}(\lambda)\, p_{XY}(a,b|\lambda) .
\end{equation}
This may be recognised as a form of Bayes theorem, where $\rho_{XY}(\lambda)\equiv p(\lambda|X,Y)$ is a probability density over the underlying variable.  
Integration is replaced by summation over discrete ranges of $\lambda$. 
The assumption of measurement independence may be formally expressed as
\begin{equation} \label{ind}
\rho_{XY}(\lambda) \equiv \rho(\lambda)
\end{equation}
for all joint settings $X$ and $Y$, i.e., the probability density is independent of the measurement settings. 

Note if $\rho_{XY}(\lambda)$ is rewritten in the conditional probablity form $p(\lambda|X,Y)$, then Eq.~(\ref{ind}) becomes $p(\lambda|X,Y)\equiv p(\lambda)$.  It follows from Bayes theorem that measurement independence is equivalent to 
\[ p(X,Y|\lambda) = p(\lambda|X,Y)\, p(X,Y)/p(\lambda) = p(X,Y) . \]
Thus the measurement settings are independent of the underlying variable $\lambda$, consistent with complete experimental freedom in choosing between them. 

The degree to which a given model satisfies measurement independence may be quantified via 
\begin{equation} \label{mdef}
M := \sup_{X,X',Y,Y'} \int d\lambda \,|\rho_{XY}(\lambda) - \rho_{X'Y'}(\lambda)| ,
\end{equation}
i.e, by the `maximum distance' between the distributions of the underlying variable for any two pairs of measurement settings.  Clearly, a distance of $M=0$ corresponds to the case of full measurement independence as per Eq.~(\ref{ind}), consistent with maximum experimental free will in choosing measurement settings.  Conversely, suppose that $M$ attains its greatest possible value, $M=2$, for some model.  Hence, there are at least two particular joint measurement settings, $(X,Y)$ and $(X',Y')$, such that for any $\lambda$ at most one of these joint settings is possible.  Hence, no experimental free will whatsoever can be exercised to choose between these settings.  Such a model has been given by Brans \cite{brans}.

The {\it fraction} of measurement independence associated with a given model may be directly quantified via
\begin{equation} \label{free}
F := 1-M/2 .
\end{equation}
Thus, $0\leq F\leq 1$, with $F=1$ corresponding to full measurement independence as per Eq.~(\ref{ind}), and $F=0$ corresponding to models having settings incompatible with any experimental free will.  

While the above definitions of $M$ and $F$ are sufficient for what follows, it is worth remarking that they may be further refined, to allow for different degrees of measurement independence for different observers.  In particular, local degrees of measurement independence, $M_1$ and $M_2$, may be defined as the distances
\[ M_1:=\sup_{X,X',Y} \int d\lambda \,|\rho_{XY}(\lambda) - \rho_{X'Y}(\lambda)| , \]
\[ M_2:=\sup_{X,Y,Y'} \int d\lambda \,|\rho_{XY}(\lambda) - \rho_{XY'}(\lambda)| . \]
Using the triangle inequality, one has the general relations $0\leq M_j \leq M\leq \min \{M_1+M_2,2\}$.  Note, for example, that $M_1=0$ implies (via Bayes theorem) that $p(X|\lambda,Y)=p(X|Y)$, i.e., the measurement setting of the first observer does not depend on the underlying variable, consistent with a free  choice of settings. Conversely, $M_1=2$ implies that settings $X$ and $X'$ exist which the first observer cannot choose between.  The models given below satisfy $M_1=M_2=M$.

One could also consider other measures of measurement independence, such as the maximum distance between $\rho_{XY}(\lambda)$ and the average underlying distribution $
\overline{\rho}(\lambda)=\int dX dY\,p(X,Y)\,\rho_{XY}(\lambda)$.  Alternatively, Barrett and Gisin have very recently suggested using the mutual information between $\lambda$ and $X$ (or $Y$) as a suitable measure \cite{barrett}.    An advantage of $M$ in Eq.~(\ref{mdef}) is that it does not depend on any of the distributions $p(X,Y)$, $p(X)$ and $p(Y)$ of measurement settings (in contrast to, for example, the maximum distance to $\overline{\rho}(\lambda)$ above, and the mutual informations in Eqs.~(2) and (6) of Ref.~\cite{barrett}).

{\it Singlet state model:}
A deterministic no-signalling model of the singlet state, having a fraction of measurement independence $F\approx 86\%$ in Eq.~(\ref{free}), will now be given.  This model generates the singlet state correlations
\begin{equation} \label{singlet} 
p_{XY}(a,b) = \frac{1}{4}(1-ab \, x\cdot y), 
\end{equation}
where $X$ and $Y$ correspond to measuring spins in directions $x$ and $y$ respectively, and $a,b\in\{-1,1\}$ denote the corresponding measurement outcomes. The requirements of no-signalling and determinism are satisfied in the model via the underlying joint probabilities in Eq.~(\ref{underlying}) having the form
\begin{equation} \label{ab}
p_{XY}(a,b|\lambda) = \delta_{a,A(x,\lambda)}\, \delta_{b,B(y,\lambda)},
\end{equation}
where the functions $A$ and $B$ take values in $\{-1,1\}$, corresponding to the deterministic measurement outcomes for measurement settings $X$ and $Y$ respectively.

The model is a modification of one first investigated by Bell \cite{bell}. In particular, as per Bell's model, the underlying variable $\lambda$ is taken to be a unit 3-vector, and the functions $A$ and $B$ are defined by
\begin{equation}
A(x,\lambda) := {\rm sign~} x\cdot\lambda,~~~~B(y,\lambda) := -\,{\rm sign~} y\cdot\lambda .
\end{equation}
However, the probability density $\rho_{XY}(\lambda)$ in Eq.~(\ref{underlying}) is defined by
\begin{eqnarray} \nonumber
\rho_{XY}(\lambda) &:=& \frac{1+x\cdot y}{8(\pi - \phi_{xy})}  {\rm ~~~for~~} {\rm sign~} x\cdot\lambda  = {\rm sign~} y\cdot\lambda,\\ \label{rho}
&:=& \frac{1-x\cdot y}{8 \phi_{xy}} {\rm ~~~for~~} {\rm sign~} x\cdot\lambda  \neq {\rm sign~} y\cdot\lambda .
\end{eqnarray}
Here, $\phi_{xy}\in [0,\pi]$ denotes the angle between measurement directions $x$ and $y$, and the density is defined to be zero when the denominators vanish.  Hence, rather than being a uniform density, $\rho_{XY}(\lambda)$ takes a first value over the regions of the unit sphere for which $\lambda$ or $-\lambda$ is within 90 degrees of both $x$ and $y$, and a second value otherwise.  

To check that the model defined by Eqs.~(\ref{ab})-(\ref{rho}) correctly reproduces the singlet state correlations (\ref{singlet}), coordinatise the unit sphere via spherical polar coordinates $(\theta,\phi)$ such that $x$ and $y$ are equatorial, with $x=(\pi/2,0)$ and $y=(\pi/2,\phi_{xy})$.  Hence, the region for which ${\rm sign~} x\cdot\lambda$ and ${\rm sign~} y\cdot\lambda$ are both equal to +1 (-1) corresponds to the spherical sector $\phi_{xy}-\pi/2 \leq \phi\leq \pi/2$ ($\pi/2+\phi_{xy}\leq \phi \leq 3\pi/2$), having area $a_+=2(\pi-\phi_{xy})$. The regions for which they are unnequal, corresponding to the second value of $\rho_{XY}$ in Eq.~(\ref{rho}), are given by the two opposing spherical sectors defined by the remaining ranges of $\phi$, each having area $a_-=2\phi_{xy}$.  Thus, Eq.~(\ref{underlying}) yields, eg,
\[ p_{XY}(+,-) = \frac{1+x\cdot y}{8(\pi - \phi_{xy})}\, a_+ = (1+x\cdot y)/4, \]
in agreement with Eq.~(\ref{singlet}).

To evaluate the degree of measurement independence for this model, note that the difference between any two densities of the form of Eq.~(\ref{rho}) will be maximised if the pair of regions corresponding to the upper value of one density has maximum overlap with the pair of regions corresponding to the lower value of the other density (and hence vice versa).  Since each pair of regions comprises two opposing spherical sectors, this maximum overlap is achieved when one pair of regions lies fully within the other pair (and hence vice versa).  Thus, if the sphere is coordinatised such that $x$ and $y$ are as per the above paragraph, complete overlap requires $x'$ and $y'$ to also lie on the equator. While there is some rotational freedom, complete overlap may then be ensured, without loss of generality, by choosing the regions to have common bisectors, so that $x'=(\pi/2,\pi/2+\phi_{xy}/2-\phi'/2)$ and $y'=(\pi/2,\pi/2+\phi_{xy}/2+\phi'/2)$ (or the antipodal points thereof). Thus, $\phi_{x'y'}=\phi'$.  Then, for example, the region for which ${\rm sign~} x'\cdot\lambda ={\rm sign~} y'\cdot\lambda =1$ corresponds to the spherical sector $\phi_{xy}/2+\phi'/2\leq \phi \leq \pi+ \phi_{xy}/2-\phi'/2$, which either encloses or is enclosed within the spherical sector for which ${\rm sign~} x\cdot\lambda = -{\rm sign~} y\cdot\lambda =-1$, i.e., $\pi/2\leq\phi\leq \pi/2+\phi_{xy}$.

The degree of measurement independence may now be calculated via Eqs.~(\ref{mdef}) and (\ref{rho}) by maximising with respect to $\phi_{xy}$ and $\phi_{x'y'}=\phi'$.  This is a straightforward but messy procedure (requiring separate consideration of $\phi_{xy}\leq\phi_{x'y'}$ and vice versa), yielding 
\begin{equation} \label{msing}
M_{singlet} = 2(\sqrt{2}-1)/3\approx 0.276 .
\end{equation}
The  maximum is achieved for $\phi_{xy}=\phi_{x'y'} =\pi/4$ (or $3\pi/4)$, which corresponds to measurement directions $x$, $y$, $x'$, $y'$ in the equatorial plane with $\phi=0,\pi/4,\pi/2,3\pi/4$, respectively, which is also the case known to maximally violate the Bell-CHSH inequality \cite{chsh}.  The same degree of measurement independence is obtained if $x'$ is replaced by $x$, implying that $M_1=M$ in this model, and hence by symmetry that $M_2=M$.

The corresponding fraction of measurement independence follows from Eqs.~(\ref{free}) and (\ref{msing}) as 
\begin{equation} \label{fsing}
F_{singlet} = (4-\sqrt{2})/3\approx 86\%
\end{equation}
The above values of $M$ and $F$ are in fact optimal for modelling the singlet state, as shown further below.

{\it Bell inequality violation:}
To investigate the extent to which Bell inequalities can be violated in general, when the assumption of measurement independence is relaxed to a given degree, let $X,X'$ and $Y,Y'$ denote possible measurement settings for a first and second observer, respectively, and label each measurement outcome by $1$ or $-1$. Further, let $\langle XY\rangle$ denote the average product of the measurement outcomes for joint measurement setting $X$ and $Y$, and define $E:=\langle XY\rangle + \langle XY'\rangle + \langle X'Y\rangle - \langle X'Y'\rangle$.  For any value $0\leq M\leq 2$, a deterministic no-signalling model is constructed below which has degree of measurement independence $M$, and for which
\begin{equation} \label{bell}
 E  = \min \{2+3M,4\}.
\end{equation}
Thus, the well known Bell-CHSH inequality \cite{chsh}, $E\leq 2$, is violated unless measurement independence is satisfied, i.e., unless $M=0$ 

It is convenient to consider the cases $M\leq 2/3$ and $M>2/3$ separately.  In the first case, consider the class of deterministic no-signalling models defined in Table~I.  These models have five underlying variables, $\lambda_1,\dots,\lambda_5$, with the outcome for measurement setting $X$ denoted by $X(\lambda_j)$, etc.  For each model these outcomes are specified by five numbers $a,b,c,d,e\in \{-1,1\}$, as shown.  The probability densities for each model, $\rho_{XY}(\lambda_j)$ etc., are defined by a single parameter, $0\leq p\leq 1/3$, as per Table~I. Summing each density over the $\lambda_j$ gives unity as required.

\begin{table}
\caption{\label{tab:table1}A class of deterministic no-signalling models violating the Bell-CHSH inequality.}
\begin{ruledtabular}
\begin{tabular}{c|cccc|cccc}
$\lambda$ &$X(\lambda)$ & $X'(\lambda)$ & $Y(\lambda)$ & $Y'(\lambda)$ & $\rho_{XY}$ & $\rho_{XY'}$ & $\rho_{X'Y}$ & $\rho_{X'Y'}$\\ 
\hline
$\lambda_1$ & $a$ & $a$ & $a$ & $a$ & $p$ & $p$ & $p$ & 0\\
$\lambda_2$ & $b$ & $-b$ & $b$ & $b$ & $p$ & $p$ & 0 & $p$\\
$\lambda_3$ & $c$ & $c$ & $c$ & $-c$ & $p$ & 0 & $p$ & $p$\\
$\lambda_4$ & $d$ & $-d$ & $-d$ & $d$ & 0 & $p$ & $p$ & $p$\\
$\lambda_5$ & $e$ & $e$ & $e$ & $e$ & $1-3p$ & $1-3p$ & $1-3p$ & $1-3p$ 
\end{tabular}
\end{ruledtabular}
\end{table}

From Table~I $\langle XY\rangle = \langle XY'\rangle = \langle X'Y\rangle = 1$, and $\langle X'Y'\rangle=1-6p$.  Hence, $E=2+6p$, and the Bell-CHSH inequality is violated by an amount $6p$.   It is also straightforward to calculate $M=2p$ and $F=1-p$, via Eqs.~(\ref{mdef}) and (\ref{free}).  Hence,  $0\leq M\leq 2/3$, and $2/3\leq F\leq 1$, with $E=2+3M$ in agreement with Eq.~(\ref{bell}).  Note that $M_1=M_2=M$ for this model.  

The choice $p=1/3$ in Table~I is of particular interest, as it demonstrates that maximum Bell inequality violation, $E=4$, can be achieved with $F=2/3$, i.e., via giving up just 1/3 of measurement independence.  In contrast, a recent theorem implies that, to achieve $E=4$, one must alternatively give up {\it all} determinism or {\it all} no-signalling (or a mixture of each) \cite{bis}.  

Note that the choice $p=1/3$ in Table~I has some unusual properties.  For example, if the underlying physical state is described by $\lambda_1$, then the joint measurement setting $X'$ and $Y'$ cannot occur, whereas there are no constraints on the other settings.  Thus, there is necessarily some degree of correlation between the two observers' measurement settings, and also between these settings and the underlying physical state.  For example, the observers can conclude, if they have performed an experiment corresponding to settings $X'$ and $Y'$, that the underlying variable is not described by $\lambda=\lambda_1$.  In this sense, it is seen that the strong correlation of measurement outcomes, as evidenced by maximal Bell inequality violation, is due in part to underlying correlations between the measurement settings - which indeed might well be expected for models in which experimental freedom of choice is restricted.

Finally, to show that there is a deterministic no-signalling model with $E=4$ for any value of $M\geq 2/3$, as implied by Eq.~(\ref{bell}), consider the modification of the class of models in Table 1 obtained by redefining the underlying probability densities as per Table~II.  It is straightforward to check that $\langle XY\rangle = \langle XY'\rangle = \langle X'Y\rangle = -\langle X'Y'\rangle = 1$, yielding $E=4$ as desired.  Moreover, the degree of measurement independence may be calculated as $M=M_1=M_2=2-4p\geq 2/3$, corresponding to $F=2p\leq 2/3$.  Note for the case $p=1/3$ that the two classes of models are equivalent.
 
\begin{table}
\caption{\label{tab:table2}A class of deterministic no-signalling models {\it maximally} violating the Bell-CHSH inequality.}
\begin{ruledtabular}
\begin{tabular}{c|cccc}
$\lambda$ & $\rho_{XY}$ & $\rho_{XY'}$ & $\rho_{X'Y}$ & $\rho_{X'Y'}$\\ 
\hline
$\lambda_1$ & $p$ & $\frac{1-p}{2}$ & $\frac{1-p}{2}$ & 0\\
$\lambda_2$ & $\frac{1-p}{2}$ & $p$ & 0 & $\frac{1-p}{2}$\\
$\lambda_3$ & $\frac{1-p}{2}$ & 0 & $p$ & $\frac{1-p}{2}$\\
$\lambda_4$  & 0 & $\frac{1-p}{2}$ & $\frac{1-p}{2}$ & $p$\\
$\lambda_5$  & 0 & 0 & 0 & 0 
\end{tabular}
\end{ruledtabular}
\end{table}

{\it Conclusions:}
In contrast to assumptions such as determinism or no-signalling, measurement independence is distinguished by not having to be completely relaxed to model the singlet state or maximum Bell inequality violation - such models require giving up only 14\% or 33\% of measurement independence respectively.  In this sense the degree of measurement dependence can be considered to be a strong `nonlocal' resource.  It would be valuable, in this context, to consider alternative measures to $M$ in Eq.~(\ref{mdef}), which admit more direct interpretations as a physical resource (eg, information-theoretic measures of correlation \cite{barrett,ce}).

It is also of interest to also consider {\it jointly} relaxing assumptions used to derive Bell inequalities.  For example, a `relaxed' Bell inequality has recently been derived, depending on the degrees to which determinism and no-signalling are relaxed \cite{bis}.  A generalisation of this inequality to include measurement independence was conjectured in Ref.~\cite{bis}, and will be proved elsewhere \cite{relaxed}.  This generalisation implies in particular that the right hand side of Eq.~(\ref{bell}), and the corresponding models in Tables~I and II, are optimal, in the sense that no larger violation of the Bell-CHSH inequality is possible for deterministic no-signalling models with a given value of $M$.  Noting that $E=2\sqrt{2}$ in Eq.~(\ref{bell}) for $M=M_{singlet}$, it similarly follows that $F_{singlet}$ represents the maximum fraction of measurement independence possible for {\it any} no-signalling deterministic model of the singlet state.

\end{document}